**Quasilinear Consequences of Turbulent Ion Heating by Magnetic Moment Breaking**

Philip A. Isenberg, Bernard J. Vasquez, and Joseph V. Hollweg

Department of Physics and Institute for the Study of Earth, Oceans and Space

University of New Hampshire, Durham NH 03824, USA

**Abstract**

The fast solar wind emerging from coronal holes is likely heated and accelerated by the dissipation of magnetohydrodynamic turbulence, but the specific kinetic mechanism resulting in the perpendicular ion heating required by observations is not understood. A promising mechanism has been proposed by Chandran et al. (2010), which in this paper we call "magnetic moment breaking" (MMB). As currently formulated, MMB dissipation operates only on the ion perpendicular motion, and does not influence their parallel temperature. Thus, the MMB mechanism acting by itself produces coronal hole proton distributions that are unstable to the ion-cyclotron anisotropy instability. This quasilinear instability is expected to operate faster than the nonlinear turbulent cascade, scattering ions into the parallel direction and generating quasi-parallel-propagating ion-cyclotron (IC) waves. To investigate the consequences of this instability on the MMB-heated protons, we construct a homogeneous model for protons with coronal hole properties. Using a simplified version of the resonant cyclotron interaction, we heat the protons by the MMB process and instantaneously scatter them to lower anisotropy while self-consistently generating parallel-propagating IC waves. We present several illustrative cases, finding that the extreme anisotropies implied by the MMB mechanism are limited to reasonable values, but the distinctive shape of the proton distribution derived by Klein & Chandran (2016) is not maintained. We also find that these combined processes can result in somewhat higher particle energization than the MMB heating alone. These quasilinear consequences should follow from any kinetic mechanism that primarily increases the perpendicular ion temperature in a collisionless plasma.

## 1. Introduction

The question of how the open solar corona and the fast solar wind are heated and accelerated continues to pose a fundamental puzzle for our understanding and modeling of the inner heliosphere. Theoretically, we know that the fast solar wind requires strong heating above the sonic critical point (Leer & Holzer 1980; Holzer & Leer 1980). Observationally, we also know that the ion population of the essentially collisionless fast wind continues to be heated as it flows outward to 1 AU (Kohl et al. 1998; Schwartz & Marsch 1983; Marsch 1991). Furthermore, in this wind emerging from coronal holes, ions are hotter than electrons and their core distributions have larger temperatures in the directions perpendicular to the local magnetic field than in the direction parallel (Marsch et al. 1982; Kohl et al. 1998; Esser et al. 1999). These perpendicularly enhanced proton distributions are often accompanied by heavy ion populations with more than mass-proportional temperatures with respect to the protons (Schmidt et al. 1980; Bochsler et al. 1985; Collier et al. 1996; Cranmer et al. 2008). Together, these facts require a heating mechanism which operates in an extended spatial region, and which preferentially heats ions perpendicular to the magnetic field with some further preference for more massive ions.

The requirement for spatially extended heating can be satisfied by invoking dissipation of an evolving turbulent cascade, driven near the Sun by either the nonlinear interaction of counter-propagating Alfvén waves (Matthaeus et al. 1999; Verdini et al. 2009; Verdini et al. 2010) or the reconnection of braided field lines (Parker 1972; van Ballegooijen 1986; Rappazzo et al. 2008), or both. The universal appearance of power-law fluctuation spectra in the solar wind has long suggested the presence of turbulent processes (Coleman 1968). Furthermore, third-moment analyses of interplanetary fluctuations have demonstrated that the turbulent energy cascade is active and can account for observed proton heating at 1 AU (Stawarz et al. 2009). The puzzle lies in specifying the kinetic process by which the turbulent dissipation may yield primarily perpendicular ion heating (Parashar et al. 2009; Markovskii & Vasquez 2011; Vasquez & Markovskii 2012; Vasquez 2015; Yang et al. 2017).

Theoretically, collisionless turbulence in a magnetized plasma is dominated by cascading fluctuations propagating perpendicular to the average magnetic field (Shebalin et al. 1983; Higdon 1984; Goldreich & Sridhar 1995), resulting in a two-dimensional (2D) power spectrum in the inertial range. This property is consistent with many observations in the solar wind (Matthaeus et al. 1990; Bieber et al. 1996; Horbury et al. 2008; Wicks et al. 2010; Chen et al. 2012). Since the fluctuation amplitudes in typical turbulent power-law spectra become very small at the high wavenumbers where dissipation takes place, it is not unreasonable to expect that linear dissipation of small-scale wave modes could provide a good description of the dissipative heating. However, quasi-perpendicular wave modes tend to dissipate through damping at the Landau resonance, heating electrons more readily than ions, and heating all particles primarily parallel to the magnetic field (Quataert 1998; Gary & Borovsky 2004). This is not what we see, so the turbulence in the solar wind must also dissipate through other means.

Several nonlinear dissipation mechanisms have been proposed to address this issue. Potentially, reconnection processes at intermittent turbulently-generated current sheets could provide a net ion heating in the perpendicular directions (Osman et al. 2011; Osman et al. 2012)). Another nonlinear mechanism relies on the disruption of ion gyromotion by the finite-amplitude fluctuations on the ion gyroscale (Chandran et al. 2010; Chandran 2010; Xia et al. 2013; Bourouaine & Chandran 2013; Klein & Chandran 2016). It is this second proposed mechanism that we focus on in this paper.

Charged particles in a magnetic field perform a circular motion in the plane perpendicular to the field. The magnetic moment, defined as $\mu = q v_\perp^2 / B$ where $q$ is the particle charge, $v_\perp$ is the perpendicular component of the particle velocity, and $B$ is the magnetic field magnitude, is an adiabatic invariant of the motion. Thus, under steady or slowly varying conditions, this quantity is conserved.[1] A particle that encounters many rapid, small-amplitude fluctuations in magnetic field intensity during a single gyration may also tend to maintain its magnetic moment, in effect averaging over the small changes in its gyro-orbit. However, if the magnetic field changes spatially on a scale near the gyroradius, or temporally on a scale near the gyrofrequency, the ordered gyromotion may be altered. Particle simulations (McChesney et al. 1987; Chen et al. 2001; Johnson & Cheng 2001) have shown that if these fluctuations in field strength at the gyrating particle's position exceed some threshold, the particle's gyromotion becomes chaotic in phase space. A distribution of particles embedded in a randomly-phased spectrum of such fluctuations would respond stochastically, with a net effect equivalent to diffusion in gyroradius. Any "thermal-like" distribution that decreases monotonically in $v_\perp$ will then have more particles moving to higher $v_\perp$ than the reverse, resulting in a net energization in the perpendicular direction.

---

[1] Since the magnetic moment is proportional to the perpendicular energy of the particle, it follows that perpendicular plasma heating must require a net increase in total ion magnetic moment. This is why gyrokinetic simulations, which rigidly assume magnetic moment conservation of all particles, cannot be used to study perpendicular heating processes.

This behavior is the basis for the formalism constructed by Chandran et al. (2010), which they termed "stochastic heating". We feel this term is ambiguous, since any heating process, in the sense of an irreversible increase in particle energy, must have a stochastic component. In this paper, we choose to call this diffusive process of Chandran et al. "magnetic moment breaking" (MMB) to distinguish it from other stochastic heating processes.

In an initially low-$\beta$ plasma (where $\beta$ is the ratio of the plasma thermal pressure to the magnetic pressure), this MMB diffusion quickly spreads the proton distribution in the perpendicular direction, potentially providing much of the perpendicular heating that would be required for the generation of the fast solar wind. In a model coronal hole, Klein & Chandran (2016) showed that the MMB heating rate can represent a substantial fraction of the turbulent dissipation rate derived for that model.

One would also like to know what the proton distribution function of an MMB-heated wind should look like, especially since the Parker Solar Probe (PSP) mission will soon provide *in situ* measurements close to the Sun (Fox et al. 2016). If the MMB mechanism results in a signature shape to the distribution, PSP observations could confirm or contradict the operation of this process. Klein & Chandran (2016) addressed this question, but they only considered the shape of a reduced distribution function, obtained by integrating their equations over the parallel proton velocity. Their reduced distribution,

$$g(v_\perp) = 2\pi \int dv_\| \, f(v_\|, v_\perp), \qquad (1)$$

was found to take on a characteristic flattop shape, described by a modified Moyal function, whose width in $v_\perp$ increased with heliocentric distance. Since the MMB formalism does not contain any particle energization in the parallel direction, this strong perpendicular heating must create proton distributions with highly anisotropic temperatures, $T_\perp/T_\parallel \gg 1$, but this property cannot be evaluated from the reduced distribution. Proton distributions with large enough thermal anisotropies are unstable to the generation of ion-cyclotron (IC) waves, propagating primarily along the magnetic field. In generating these waves, the protons scatter to a more isotropic state, reducing the free energy in the anisotropy.

In this paper, we explore this issue with a simplified "toy" model to show the likely interplay between the MMB heating and the quasilinear response of the protons. We consider a homogeneous plasma of protons and massless electrons interacting with a steady power-law spectrum of 2D turbulence. We describe the simplified computational model in the next section. Section 3 presents the results of three illustrative examples, while Section 4 contains our discussion and conclusions.

## 2. Simplified Computational Model

*a) MMB heating*

We first outline the proton response to MMB heating by a steady power-law spectrum of 2D turbulence in a spatially homogeneous system. Here, we follow the formalism presented by Chandran et al. (2010) and Klein & Chandran (2016). Specifically, we solve the equation for the proton distribution function, $f(v_\parallel, v_\perp, t)$,

$$\frac{df}{dt} = \frac{1}{v_\perp} \frac{\partial}{\partial v_\perp} \left[ v_\perp D_{MMB} \frac{\partial f}{\partial v_\perp} \right], \tag{2}$$

where the diffusion coefficient takes the form

$$D_{MMB}(v_\perp) = c_1 \, \Omega_p \frac{\left(\delta v_\rho\right)^3}{v_\perp} \exp\left(-c_2 \frac{v_\perp}{\delta v_\rho}\right). \tag{3}$$

The quantity $\delta v_\rho (k_\perp)$ is the amplitude of the turbulent $\mathbf{E} \times \mathbf{B}$ plasma velocity at the perpendicular scale of a proton gyroradius, $\rho$, given by the spectral wavenumber $k_\perp = \Omega_p/v_\perp$. Here, the proton gyrofrequency is $\Omega_p = eB/mc$, $e$ and $m$ are the proton charge and mass, and $c$ is the speed of light. The cubic factor in expression (3) estimates the diffusion of the proton's gyroradius due to the magnetic moment disruption by electric fields in the turbulent background that vary on the scale of its cyclotron orbit. The exponential factor models the suppression of this diffusion when the fluctuation electric fields are small and are effectively averaged over by the particle's gyromotion. The parameters $c_1$ and $c_2$ are constants of order unity to be inferred from simulations and observations.

In this model, we take the 2D turbulent spectrum to have a power-law inertial range with index $-3/2$ (Boldyrev 2006; Boldyrev et al. 2011; Perez et al. 2012), so

$$\left\langle \delta v^2 \right\rangle = \left(\delta v_o\right)^2 \left(\frac{k_\perp}{k_o}\right)^{-1/2}, \qquad (4)$$

where $\delta v_o$ is the amplitude of the turbulent perpendicular velocity at the wavenumber $k_o$, representing the outer scale of the turbulence. We assume this spectrum steepens to $k_\perp^{-3}$ at the gyroradius scale for thermal protons, corresponding to a dissipation range. It is useful to normalize all velocities to the local Alfvén speed, $V_A = B/\sqrt{4\pi mn}$ where $n$ is the local proton number density. In this case, this spectrum yields an MMB diffusion coefficient

$$D_{MMB}(v_\perp) \sim \begin{cases} \left(\dfrac{v_\perp}{V_A}\right)^{-1/4} \exp\left[-d_2\left(\dfrac{v_\perp}{V_A}\right)^{3/4}\right] & \dfrac{v_\perp}{V_A} > \sqrt{\beta_o} \\[4mm] \beta_o^{-9/8}\left(\dfrac{v_\perp}{V_A}\right)^2 \exp\left[-d_2\beta_o^{3/4}\right] & \dfrac{v_\perp}{V_A} < \sqrt{\beta_o} \end{cases} \qquad (5)$$

where

$$d_2 = c_2\left(\frac{V_A}{\delta v_o}\right)\left(\frac{\Omega_p}{k_o V_A}\right)^{1/4} \qquad (6)$$

and $\beta_o$ is the initial value of the plasma $\beta$. Formally, the remaining proportionality constant is absorbed into the time variable of equation (2), giving an MMB time scale $\tau_{\mathrm{MMB}} \equiv d_1\, t$, where

$$d_1 = c_1 \Omega_p \left( \frac{\delta \upsilon_o}{V_A} \right)^3 \left( \frac{k_o V_A}{\Omega_p} \right)^{3/4} . \tag{7}$$

To illustrate the effect of this heating process on its own, we solve equation (2) for $f(\upsilon_{\parallel}, \upsilon_{\perp}, \tau)$, starting with a Maxwellian distribution of protons with $\beta_o = 10^{-4}$, such as might be expected in the solar wind generation region of a coronal hole. The value of $c_2$ in the exponential term is taken here as $c_2 = 0.15$, from Chandran (2010), though values from 0.17 - 0.44 have also been studied. We choose the turbulent amplitude, $\delta \upsilon_o / V_A = 0.0573$, and the normalized outer scale, $k_o V_A / \Omega = 7 \times 10^{-5}$. We construct a rectangular grid in the cylindrical $\upsilon_{\parallel}$ - $\upsilon_{\perp}$ space, assuming symmetry in $\upsilon_{\parallel}$. In our computational model, all velocities are normalized to the Alfvén speed. In this case, the grid consists of 300 points between $0 \leq \upsilon_{\parallel}/V_A \leq 0.15$, and 1000 points between $0 \leq \upsilon_{\perp}/V_A \leq 0.5$. To implement the reflecting boundaries at $\upsilon_{\parallel} = 0$ and $\upsilon_{\perp} = 0$ we stagger the grid, shifting it upward by one-half of the spacing in both $\upsilon_{\parallel}$ and $\upsilon_{\perp}$. We take the outer boundaries, now at $\upsilon_{\parallel}/V_A = 0.1505$ and $\upsilon_{\perp}/V_A = 0.5005$ to be absorbing, so $f = 0$ there. The time evolution of the proton distribution is obtained on this grid by a standard Crank-Nicholson implicit scheme with a step size of $\Delta t = 20(\Delta v)^2$, where $\Delta v$ is the grid spacing in $\upsilon_{\perp}/V_A$.

We integrate equation (2) for $3.6 \times 10^6 \Delta t$, and the results are shown in Figure 1. Figures 1a and 1b show color spectrograms of the proton distribution function at the start and end times, each normalized to their maximum value at the phase-space origin. In Figure 1c, we show the time development of the perpendicular temperature, defined in the usual way as the second moment of the distribution in $\upsilon_{\perp}$, normalized to its initial value. We see that the protons are heated rapidly in the perpendicular direction, as expected from this mechanism. Since the parallel temperature does not change under this formalism of MMB heating, these distributions correspond to thermal anisotropies which also grow continuously, well past $T_{\perp}/T_{\parallel} = 300$. The heated distributions are broadly extended in $\upsilon_{\perp}$ until they fall off sharply due to the exponential cutoff in the MMB diffusion coefficient.

Information on the thermal anisotropy is lost when the distributions are integrated over $v_\parallel$. The resulting reduced distributions (1) of our MMB-heated cases have structures similar to those obtained by Klein & Chandran (2016), as shown in Figures 5, 6e, and 8e below. We note that our simple homogeneous model differs from that of Klein & Chandran, in that it does not include a perpendicular cooling corresponding to the radial expansion of the solar wind. Nevertheless, the reduced distributions in both cases exhibit the same characteristic flattop core and sharp cutoff, as expected from the structure of the perpendicular diffusion coefficient (5).

Of course, thermal anisotropies of the magnitude shown in Figure 1c are not sustainable since such distributions are unstable to the generation of cyclotron-resonant IC waves. The quasilinear wave-particle interaction efficiently reduces the particle anisotropy, scattering highly perpendicular protons into the parallel direction. The energy lost by the perpendicular protons appears in quasi-parallel IC waves, representing a channel for the transfer of turbulent fluctuations cascading in $k_\perp$ into wave power at high $k_\parallel$. In this paper, we consider the quasilinear evolution of the protons and waves under the combined effects of turbulent MMB heating and resonant cyclotron scattering, using a simplified treatment of the wave-particle interaction described in the next subsection.

*b) Quasilinear proton anisotropy instability and wave generation*

In a low-$\beta$ proton-electron plasma, the quasilinear cyclotron resonant interaction is a straightforward consequence of two physical properties: the proton cyclotron resonance and the particle energy conservation in the frame of the resonant wave.

The electric field of a parallel-propagating IC wave rotates about the large-scale magnetic field in the same direction as the proton's gyromotion. When the Doppler-shifted frequency of the wave as seen by a streaming proton is equal to the proton's gyrofrequency, the wave and particle are in resonance and can exchange energy efficiently. This condition for resonance is

$$\omega(k) - k_\parallel \, v_\parallel \; = \; \Omega_p \; , \qquad (8)$$

where $\omega$ is understood as the real part of the generalized complex wave frequency, $\omega =$ Re($\varpi$). Since total energy is conserved, waves will be generated (Im($\varpi$) > 0) if the protons resonant with these waves lose energy in the interaction.[2]

During this resonant energy exchange, a proton will conserve its energy as measured in the reference frame moving along the large-scale magnetic field with the parallel phase speed of the resonant wave, $\omega /k_{\parallel}$. This property follows from the fact that the magnetic field of the wave is independent of time in that reference frame, so the electric field required for energy change in that frame is transformed away. Thus, the self-consistent proton response to the resonant interaction is a scattering in phase space along a surface which is locally a sphere centered on the point $(v_{\parallel},\ v_{\perp}) = (\omega /k_{\parallel}, 0)$. The structure of these resonant surfaces depends on the wave dispersion relation, $\omega (k_{\parallel})$, where $k_{\parallel}$ is coupled to $v_{\parallel}$ through (8).

In a many-particle system interacting with a broad spectrum of randomly-phased waves, this particle motion along the resonant surfaces will be diffusive. Thus the particles will undergo a net transport down any phase-space gradient along the resonant surface, given by the standard quasilinear expression

$$G \equiv v_{\perp}\frac{\partial f}{\partial v_{\parallel}} + \left(\frac{\omega}{k} - v_{\parallel}\right)\frac{\partial f}{\partial v_{\perp}}\ . \qquad (9)$$

(Kennel & Engelmann 1966; Rowlands et al. 1966; Gendrin 1968; Lee 1971; Isenberg & Lee 1996). If, at some $v_{\parallel}$, this transport yields a net loss of proton energy as measured in the plasma frame, the waves resonant with those protons will grow. The growth rate at the resonant $k_{\parallel}$ will be proportional to the $v_{\perp}$-integral of this resonant gradient, evaluated at constant $v_{\parallel}$. Furthermore, if this interaction proceeds in time under otherwise constant conditions, the gradient (9) will tend towards zero, yielding a proton distribution that is constant along the part of each resonant surface that was once unstable to wave growth. The case where $G = 0$ throughout the proton distribution defines the state of marginal stability, Im ($\varpi$) = 0, in principle the end point of the quasilinear interaction.

---

[2] In this paper, we choose the convention that $\omega \geq 0$ for IC waves, and their propagation direction is set by the sign of $k_{\parallel}$. This is a different convention from that chosen in some other papers, such as Isenberg & Lee (1996).

These aspects of the quasilinear resonant interaction are fairly general, but they are most useful when the resonance relationship is simple. The low-$\beta$ anisotropy instability is dominated by the cyclotron resonant interaction between the protons and parallel-propagating IC waves. In this illustrative model, we take the dispersion relation for these waves to be that found in a cold plasma, which has the analytical form in the rest frame of the plasma

$$
\begin{aligned}
\omega_\pm(k) &= \pm k_\parallel V_A \sqrt{1 - \frac{\omega}{\Omega_p}} \\
&= \pm \frac{k_\parallel V_A}{2} \left( \sqrt{y^2 + 4} \mp y \right)
\end{aligned}
\tag{10}
$$

where $y = k_\parallel\, V_A / \Omega_p$ and the +(−) refers to waves propagating forward (backward) with respect to the magnetic field, designated by $k_\parallel > (<)\, 0$.

We note that IC dispersion relations are often more complicated than that of equation (10). The presence of minor ions, if they are cold enough, causes additional branches for $\omega(k_\parallel) < \Omega_p$ (e.g. (Isenberg 1984)). Even a single-branched dispersion relation as used here can yield multiple resonances with heavy ions, causing a distinctive overlap of heavy ion resonant surfaces (Isenberg 2001; Isenberg & Vasquez 2007, 2009; Isenberg et al. 2010). Furthermore, protons streaming sufficiently faster than $V_A$ with respect to the bulk plasma may also resonate with parallel fast-mode waves (Isenberg & Lee 1996; Isenberg 2005), though this additional interaction may be neglected in the low-$\beta$ conditions of the inner heliosphere.

In this paper, we restrict ourselves to the simplest case where these complications do not arise. In our simplified model, we also assume the waves are described by (10) at all times, so the self-consistent effects of the MMB proton heating on the dispersion relation are not treated. We have previously investigated the self-consistent solution of the IC dispersion relation for a thermal proton-electron plasma at marginal stability (Isenberg 2012; Isenberg et al. 2013; see also Cranmer 2014) and found that the qualitative behavior of the wave-particle interaction – anisotropic protons scattering down the quasilinear resonant surfaces, reducing the anisotropy and generating waves – is similar to that presented here.

The computation of the resonant surfaces for this interaction starts from the simultaneous solution of the resonance condition (8) and the dispersion relation (10). The character of this solution can be seen from a plot of the two expressions in the same $\omega$-$k_\parallel$ plane, as shown in Figure 2. The resonance condition defines a straight line in this plane, with slope $v_\parallel$ and intercept at $(\omega, k_\parallel) = (\Omega_p, 0)$. For every proton, the phase speed of its resonant wave is specified by the point where this straight line intersects the dispersion curve. Figure 2 shows an example of resonance for a backward streaming proton ($v_\parallel < 0$) with a forward-propagating ion cyclotron wave ($k_\parallel > 0$) in the plasma rest frame. These relations are symmetric about $k_\parallel = 0$, so forward streaming protons will resonate with backward-propagating waves (not shown).

The quasilinear resonant surfaces corresponding to a particular interaction are then obtained from the condition of energy conservation in the reference frame of the wave. Specifically, at each $v_\parallel$ the surface will locally follow a sphere centered about the point $(v_\parallel, v_\perp) = (V_r, 0)$, where we label $V_r (v_\parallel)$ as the phase speed of the wave resonant with the protons streaming at $v_\parallel$ (Isenberg & Lee 1996). Integrating this condition defines a family of surfaces as

$$v_\parallel{}^2 + v_\perp{}^2 - \int_0^{v_\parallel} V_r(v_\parallel{}')dv_\parallel{}' = \eta^2, \tag{11}$$

where $\eta$ is an integration constant which labels each surface. In equation (11), we have set $\eta$ equal to the value of $v_\perp$ where a given surface intersects the $v_\parallel = 0$ axis. One can see from Figure 2 that the simple system used here describes a one-to-one relationship between the proton streaming speed, $v_\parallel$, and the wavenumber and phase speed of the resonant IC wave. In this case, the resonant surfaces form a nested set, and do not intersect one another. The analytic dispersion relation for parallel-propagating IC waves (10) allows a closed form description of these IC resonant surfaces in terms of the resonant wavenumber (Isenberg & Lee 1996)

$$v_\perp{}^2 = \eta^2 - V_A{}^2 \left[ \frac{1}{y^2} - \ln\left( \frac{y \pm \sqrt{y^2 + 4}}{2y} \right) \right] \tag{12}$$

$$v_{\parallel} = V_A \left( \frac{-1}{y} \pm \frac{\sqrt{y^2 + 4} \mp y}{2} \right). \tag{13}$$

We wish to apply these concepts of resonant wave-particle scattering to estimate the effects of the quasilinear resonant interaction on proton distributions heated by the MMB mechanism. In this sense, we expect that nonlinear processes, such as MMB heating or the turbulent cascade itself, will proceed on slower timescales than a quasilinear process like the resonant cyclotron interaction. In a theoretical description, the changes due to these processes come about through multiple products of small-amplitude quantities. Nonlinear effects typically appear in the kinetic particle equations at higher order in these small quantities than quasilinear interactions, so the quasilinear evolution tends to occur faster. Our simple model makes the assumption that the unstable portion of a proton distribution will relax to the state of marginal stability in an essentially infinitesimal time compared to the nonlinear MMB heating.

We incorporate this rapid quasilinear relaxation into the computational model of §2a by periodically redistributing the proton density on the unstable portion of the phase-space grid to enforce $G = 0$ there. Specifically, as the MMB heating increases the temperature anisotropy in time, we test the cumulative value of the gradient (9) and determine the range of $v_{\parallel}$ where the growth rate

$$\text{Im}(\varpi) \sim \int_0^\infty G\left(v_{\parallel}, v_{\perp}\right) v_{\perp}^2 \, dv_{\perp} > 0 \, . \tag{14}$$

At each grid point in this unstable region, we construct the resonant surface passing through that point and compute the interpolated values of $f$ along that surface. For each resonant surface, we then find the maximum value of $f$ and average the density on the portion of that surface with smaller $v_{\perp}$ than the position of that maximum. If the grid point in question is in this lower portion of the resonant surface, we replace the density there with this average value. We limit this averaging to the lower portion of the resonant surface since diffusion down the density gradient in that direction will result in lower particle energy, which is then transferred to the unstable waves. Conversely, scattering particles into the upper portion of the resonant surface would produce an increase of particle energy, so would not be a valid response to instability.

Following each quasilinear redistribution of particles, we obtain the new waves generated by the diffusive particle transport from energy conservation, using the method of Johnstone et al. (Johnstone et al. 1991; Huddleston & Johnstone 1992). The differential energy lost by each particle in the unstable range of $\upsilon_{\parallel}$ as it scatters through $\upsilon_{\parallel} + d\upsilon_{\parallel}$ along a resonant surface is given by (Isenberg & Lee 1996)

$$\frac{d\varepsilon}{d\upsilon_{\parallel}} = m\left(\upsilon_{\parallel} + \upsilon_{\perp}\frac{d\upsilon_{\perp}}{d\upsilon_{\parallel}}\right)\Bigg|_{\eta=\text{const.}} = mV_r(\upsilon_{\parallel}) \tag{15}$$

where $\varepsilon = m\,(\upsilon_{\parallel}{}^2 + \upsilon_{\perp}{}^2)/2$, and the second step follows from (11). Writing this energy change as a function of the normalized resonant wavenumber yields, from (10) and (13),

$$\frac{d\varepsilon}{dy} = \frac{d\varepsilon}{d\upsilon_{\parallel}}\frac{d\upsilon_{\parallel}}{dy} = \frac{m}{2}\left(y - \frac{1}{y} + \frac{y^4 + y^2 - 4}{y^2\sqrt{y^2 + 4}}\right) \tag{16}$$

We note that, for $\upsilon_{\parallel} > 0$, we have $y < 0$ and (16) is a negative quantity. The number of particles scattered through each $\upsilon_{\parallel}$ grid line corresponds to the density increase beyond that grid line due to the redistribution. This is simply

$$\Delta n(\upsilon_{\parallel}) = 2\pi\int_{\upsilon_{\parallel}}^{\infty} d\upsilon_{\parallel}{}'\int_{0}^{\infty}\left(f_{QL} - f_{MMB}\right)\upsilon_{\perp}\,d\upsilon_{\perp} \tag{17}$$

where $f_{MMB}$ and $f_{QL}$ are the proton phase space densities before and after the quasilinear redistribution, respectively. Thus, the wave energy spectrum after each redistribution step will be incremented by

$$\Delta E_w(y) = -\frac{d\varepsilon}{dy}\Delta n(\upsilon_{\parallel}) \tag{18}$$

where the wavenumber $y$ is related to the parallel velocity $\upsilon_{\parallel}$ through (13).

To display the IC spectral enhancements in a standard way, we obtain the fluctuating magnetic field intensity, which is a particular fraction of the total wave energy as a function of wavenumber. For the cold plasma dispersion relation (10) used here, this fraction is given by

$$\frac{\left\langle\delta B^2(k)\right\rangle}{8\pi} = \frac{1 - \omega(k)/\Omega_p}{2 - \omega(k)/\Omega_p}E_w(k)\,. \tag{19}$$

As we have not been able to find a derivation of this relationship in the literature, we provide one in the Appendix.

In this system, the anisotropy instability is driven by MMB perpendicular heating of an initially Maxwellian proton distribution. We will see that the unstable range of $v_\parallel$ starts at the high-$v_\parallel$ edge of the distribution, with a lower limit that quickly progresses to smaller $v_\parallel$ at increasing times. We follow this evolution in the initial stages of the heating, but our finite grid does not allow us to track the lower limit after it falls below the first $v_\parallel$-grid line, continuing towards $v_\parallel = 0$. However, we expect that small-scale plasma fluctuations would maintain fairly smooth particle distributions for very small $v_\parallel$. In these computations, we will take the unstable region to extend to $v_\parallel = 0$ ($k_\parallel \rightarrow \infty$) once the proton distribution at the first $v_\parallel > 0$ grid line has satisfied (14).

*c) Illustrative examples*

We will apply this simplified model of MMB heating and quasilinear wave generation to three examples with a range of initial values for $\beta_o$. In each case, we will choose the values of the other physical parameters to suggest different radial regions in a coronal hole.

The three cases are labeled A, B, and C, and are initialized with $\beta_o$ equal to $10^{-4}$, $10^{-3}$, and $10^{-2}$, respectively. We will take the other variables to correspond to physical conditions in the model coronal hole of Isenberg & Vasquez (2011) near heliocentric radial positions of $r = 2\,R_s$, $6\,R_s$, and $15\,R_s$, for the respective cases. That coronal hole model considered a narrow radial flux tube with a super-radial expansion, given by the area function

$$A(r) = \frac{5}{16} \frac{R^6}{R^4 + 4} \qquad (20)$$

where $R = r/R_s$. We will take the amplitude of the turbulent fluctuations at the outer scale consistent with the work of Cranmer & van Ballegooijen (2005), and outer scale wavenumber to be set at the width of the model flux tube, $k_o = 1.4 \times 10^{-4}/\sqrt{A}$ km$^{-1}$. The choice of Alfvén speed for each case, with the density profile from Isenberg & Vasquez (2009), determines the background magnetic field intensity and the value for the proton gyrofrequency. With these physical values, we can also evaluate the MMB time scale $\tau_{MMB} = 1/d_1$ from equation (7), where we set the parameter $c_1 = 1$. The values of these quantities for each case are given in Table 1.

The computational system here is symmetric in $v_\parallel$ and $k_\parallel$. In our presentation and discussion of the results, we will refer to both of these variables as positive with the understanding that the negative half of the particle distributions and the wave spectra behave identically to the ones shown.

The computational grids are staggered rectangular grids of 1000 x 1023 points in $v_\parallel$ - $v_\perp$ space, where the speeds are normalized to $V_A$, and the unstaggered outer boundaries in phase-space are set to $v_{max}/V_A = 0.5$, 2, and 4 for both variables, respectively in each case. We retain the computational time step of $\Delta t = 20(\Delta v)^2$ for the MMB heating in all cases. The quasilinear redistribution is much more computationally intensive, so we redistribute the protons and increment the IC wave spectrum only every 10 of these MMB steps. For each computational case, the total proton density was conserved to within 0.3%.

## 3. Results

The combined quasilinear proton and wave computation for Case A was run for $6.5 \times 10^5$ time steps, corresponding to a period of more than 600 seconds using the physical parameter values for 2 $R_s$ given in Table 1. The initial proton distribution for Case A was the same as shown in Figure 1a. The MMB heating rapidly increased the perpendicular temperature and the anisotropy, leading to unstable conditions defined by equation (14). The unstable portions of the proton distribution were treated as described in §2b, redistributing particles to impose $G = 0$ there and incrementing the wave intensities accordingly.

For this case, the unstable region appeared first at the high-$v_\parallel$ edge of the proton distribution and progressed downward in $v_\parallel$ for the first 970 steps (< 1 s) until the entire distribution was contributing to the wave growth as it was heated. Figure 3b shows the normalized proton distribution at 600 s from this combined model, while Figure 3a shows the normalized distribution at the same time for MMB heating only. The quasilinear pitch-angle scattering that accompanies the wave growth broadens the proton distribution substantially. Figure 3c plots the evolution of the proton temperature components, along with the perpendicular temperature under the MMB heating only from Figure 1c. It is clear that the quasilinear response to the anisotropy instability in this model has transferred some of the proton thermal energy due to MMB heating from the perpendicular directions into the parallel direction.

Figure 3d shows the evolution of the thermal anisotropy for the combined model. In contrast with the MMB-only result of Figure 1b, the quasilinear wave-particle interaction for this case has reduced the anisotropy to much more reasonable levels. We note that the total proton energy in the quasilinear computation seems larger than that resulting from the MMB-only case, and we address this energy behavior below.

Figure 4 shows the corresponding IC wave spectrum in Case A for three times during the evolution, using (18) and (19). The wave intensities here are small, with a growing peak at wavenumbers resonant with the thermal protons.

Finally, we note that the addition of quasilinear resonant scattering to the proton evolution produces distributions considerably different from those predicted by Klein & Chandran (2016) for MMB heating alone. Figure 5 shows the reduced proton distributions of Figures 3a and 3b, comparing the results of our model with MMB only heating after 600 s. The characteristic flattop shape exhibited by the reduced distributions of Klein & Chandran has been replaced by a less distinctive rounded shape. We suggest that these quasilinear effects will substantially erase the particular indicators of MMB heating derived by Klein & Chandran, so that the definitive observational test of this mechanism proposed in that paper may not be so straightforward.

Cases B and C start with larger values of $\beta$ and are intended to represent coronal hole conditions farther from the Sun. We find these results to be qualitatively similar to those of Case A. Case B was run for $3 \times 10^5$ time steps, corresponding to more than 500 s under the physical conditions near 6 $R_s$. Case C was run for $2 \times 10^5$ time steps, equal to 1500 s near 15 $R_s$. Figure 6 displays the particle results for Case B in the same format as Figure 3, showing the normalized proton distributions at 500 s for the MMB only (a) and quasilinear (b) computations, the time evolution of the temperatures (c), and the anisotropy (d). Figure 6e shows the comparison of the reduced distributions for the results of Figures 6a and 6b. Figure 7 shows the self-generated IC wave spectra for Case B at three times in the computation.

Figures 8 and 9 give the same results for Case C at 1500 s. We see that broader initial distributions eventually lead to smaller anisotropies and larger IC wave intensities. For all cases, the reduced distribution functions (1) are rounded, no longer displaying the distinctive flattop shapes of Klein & Chandran (2016).

As mentioned above, we note from the temperature plots in Figures 3c, 6c, and 8c that the quasilinear extension of MMB heating in this model appears to yield somewhat more heating to the protons than the MMB heating alone provides. This is shown in Figure 10, where we compare the total proton temperature, $T_{\text{tot}} = (2\,T_\perp + T_{||})/3$, for the quasilinear model (red, solid line) to the MMB-only result (blue, dashed line) in the three cases. The broader distributions evolving from hotter initial protons are energized to a greater extent by the quasilinear scattering. At the final times shown, the ratios of the quasilinear temperatures to the MMB-only temperatures are 1.074, 1.173, and 1.233 for the Cases A, B, and C, respectively.

We attribute this additional energization to a recycling effect resulting from the pitch-angle scattering. When the quasilinear interaction scatters heated perpendicular protons to larger $v_{||}$, it also transports them to smaller $v_\perp$. These particles can then participate in further rounds of MMB heating, taking on more energy than would be the case for the MMB process alone. In our simple model, the available energy for these multiple scatterings is not limited, since we take the background turbulent intensity to be constant. A more detailed model would be necessary to determine the importance of this effect in a realistic coronal hole.

## 4. Discussion and Conclusions

The acceleration and heating of the solar wind is likely driven by dissipation of MHD turbulence in the corona. In coronal holes, processes driving the fast solar wind preferentially heat ions in the directions perpendicular to the large-scale magnetic field. One such turbulent dissipation process is what we call "magnetic moment breaking", as proposed by Chandran et al. (2010). As currently formulated, MMB heating operates only on the ion perpendicular motion. Thus, this process acting on its own will inevitably lead to highly anisotropic ion distributions with $T_\perp/T_\parallel \gg 1$. However, it is well known that perpendicular anisotropies are strongly limited by quasilinear kinetic instabilities. In the low $\beta$ plasma of the solar corona, the resonant cyclotron interaction will efficiently transfer perpendicular ion energy in the parallel direction, generating primarily parallel-propagating ion cyclotron waves as a result.

In this paper, we have constructed a simplified "toy" model to investigate the quasilinear evolution of a proton distribution when it is heated by the MMB mechanism. We have found that this quasilinear interaction significantly broadens the proton distribution in the parallel direction, limiting the thermal anisotropy to small values. The reduced distributions, obtained by integrating over the parallel phase-space velocity of the particles, are no longer described by the modified Moyal function of Klein & Chandran (2016), but are rounded in a less distinctive manner.

This unstable interaction also generates a growing spectrum of parallel-propagating IC waves at thermal wavenumbers, $k_\parallel V_A/\Omega_p \sim 1$. Our simple model does not suggest how these new fluctuations might evolve in an inhomogeneous coronal hole, or whether they might become an important ingredient in further kinetic developments. Eventually, they might also contribute to the "slab" component of observed turbulence in the *in situ* solar wind. In any case, since these waves are a direct consequence of the perpendicular proton heating assumed to be driving the solar wind, this model implies that the intensity of these waves should be larger at smaller distances from the Sun. Thus, we predict that the instruments on the Parker Solar Probe will detect increased intensities of quasi-parallel-propagating IC waves as it travels closer to the Sun.

This simple model is unable to address the source of suprathermal tails in the solar wind particle distributions. The only energy source included here is the specific form of MMB diffusion described by (2) and (3). In this form, the exponential cutoff in the diffusion coefficient (3) prevents the development of power-law solutions characteristic of observed suprathermals. The quasilinear particle redistribution itself does not provide further energization in this model. Thus, if suprathermal particles are generated in the solar wind acceleration region, they are unlikely to result from this form of turbulent dissipation without the operation of additional processes.

We emphasize that the effects of quasilinear scattering as exhibited in the model of this paper are not specific to the MMB heating mechanism. This quasilinear wave-particle interaction is expected to operate on any sufficiently perpendicular proton distribution in a similar way. Thus, any kinetic heating mechanism which preferentially heats ions in the perpendicular directions enough to drive the fast solar wind is expected to trigger this resonant cyclotron anisotropy instability, broadening the particle distributions, limiting the anisotropies, and generating quasi-parallel-propagating IC waves qualitatively as discussed here.

In subsequent work, we will use the proton distributions from this simplified model to initiate more detailed computational investigations of the kinetic effects of turbulent MMB heating in a coronal hole.

## Appendix

Energy partition of parallel-propagating ion cyclotron waves

In this Appendix, we derive the expression (19) for the fraction of total wave energy contained in the magnetic field fluctuations for a parallel-propagating ion cyclotron (IC) wave in a cold proton-electron plasma. The method of Johnstone et al. (1991) to obtain resonant wave intensities from the changes in an unstable particle distribution depends on the application of energy conservation. In this technique, the energy losses experienced by the scattering particles are assumed to appear in the resonant waves, which yields a spectrum for the total fluctuating energy. In most previous applications of this technique, the resulting magnetic intensity spectrum was obtained under the further assumption of equipartition between the fluctuating particle kinetic energy and the fluctuating magnetic field energy, as would be the case for Alfvén waves. However, the computations here explicitly include the IC wave dispersion, and the self-generated wave spectra need to take dispersive effects into account.

We start with the basic electromagnetic equations of Faraday's Law

$$\nabla \times \mathbf{E} = -\frac{1}{c}\frac{\partial \mathbf{B}}{\partial t} \tag{A1}$$

and the electromagnetic force on a proton

$$m\frac{\partial \mathbf{v}}{\partial t} = e\left(\mathbf{E} + \frac{\mathbf{v}}{c}\times\mathbf{B}\right). \tag{A2}$$

Assuming a steady background magnetic field, $\mathbf{B_o}$, in the $\mathbf{z}$ direction, and a wave propagating only along that field, we Fourier transform the fluctuating quantities so that each is equal to a Fourier amplitude ($\delta\mathbf{v}$, $\delta\mathbf{B}$, $\mathbf{E}$) times the function $\exp[i(k\,z - \omega t)]$. Keeping terms to the first order in the fluctuations only, these equations then become

$$k\hat{\mathbf{z}}\times\mathbf{E} = \frac{\omega}{c}\delta\mathbf{B} \qquad \text{and} \qquad -im\omega\delta\mathbf{v} = e\left(\mathbf{E} + \frac{\delta\mathbf{v}}{c}\times\mathbf{B_o}\right). \tag{A3}$$

The Fourier amplitudes only have $(x, y)$ components, and these can be combined into a helical format $(...)_\pm = [(...)_x \pm i\,(...)_y]/2$ to yield

$$E_\pm = \mp i\frac{\omega}{ck}\delta B_\pm \qquad \text{and} \qquad \frac{e}{m}E_\pm = -i\left(\omega \mp \Omega_p\right)\delta v_\pm, \tag{A4}$$

where $\Omega_p = qB_o/mc$ is the usual proton cyclotron frequency. It follows that

$$\delta v_\pm = \pm\frac{\Omega_p}{\omega \mp \Omega_p}\frac{\omega}{k}\frac{\delta B_\pm}{B_o} \tag{A5}$$

for these linear parallel-propagating waves.

The total wave energy is defined as

$$\mathcal{E}_\pm = \frac{1}{2}\rho\left(\delta v_\pm\right)^2 + \frac{1}{8\pi}\left(\delta B_\pm\right)^2 = \left[1 + \left(\frac{\Omega_p}{\omega \mp \Omega_p}\right)^2\left(\frac{\omega}{kV_A}\right)^2\right]\frac{\left(\delta B_\pm\right)^2}{8\pi}. \tag{A6}$$

Parallel-propagating IC waves are polarized in the left-hand sense, corresponding to the subscripted plus sign in the helical representation. Inserting the cold-plasma dispersion relation (10) to describe the waves in this paper, we find that the fluctuating magnetic intensity for these IC waves is given by the fraction

$$\frac{\left(\delta B_+\right)^2}{8\pi} = \frac{1 - \omega/\Omega_p}{2 - \omega/\Omega_p}\mathcal{E}_+ \tag{A7}$$

as shown in (19).

**Acknowledgements.** The authors are grateful for valuable conversations with B. D. G. Chandran, T. G. Forbes, K. G. Klein, M. A. Lee and D. Verscharen. Many computations were performed on Trillian, a Cray XE6m-200 supercomputer at UNH supported by the NSF MRI program under grant PHY-1229408. This work was also supported in part by NASA grants 80NSSC17K0009, and 80NSSC18K1215.

Table 1. Physical values taken for each illustrative case.

| Case | $\beta_o$ | $R\ (R_s)$ | $V_A$ (km/s) | $\delta\upsilon_o$ (km/s) | $\delta\upsilon_o/V_A$ | $k_o$ (km$^{-1}$) | $\Omega_p$ (s$^{-1}$) | $\tau_{\mathrm{MMB}}$ (s) |
|------|-----------|------------|--------------|---------------------------|------------------------|-------------------|-----------------------|---------------------------|
| A | $10^{-4}$ | 2 | 3000 | 172 | 0.0573 | $1.4\times10^{-4}$ | 5978 | 185. |
| B | $10^{-3}$ | 6 | 1000 | 263 | 0.263 | $4.18\times10^{-5}$ | 313.3 | 22.5 |
| C | $10^{-2}$ | 15 | 500 | 221 | 0.442 | $1.67\times10^{-5}$ | 69.46 | 23.1 |

**Figures**

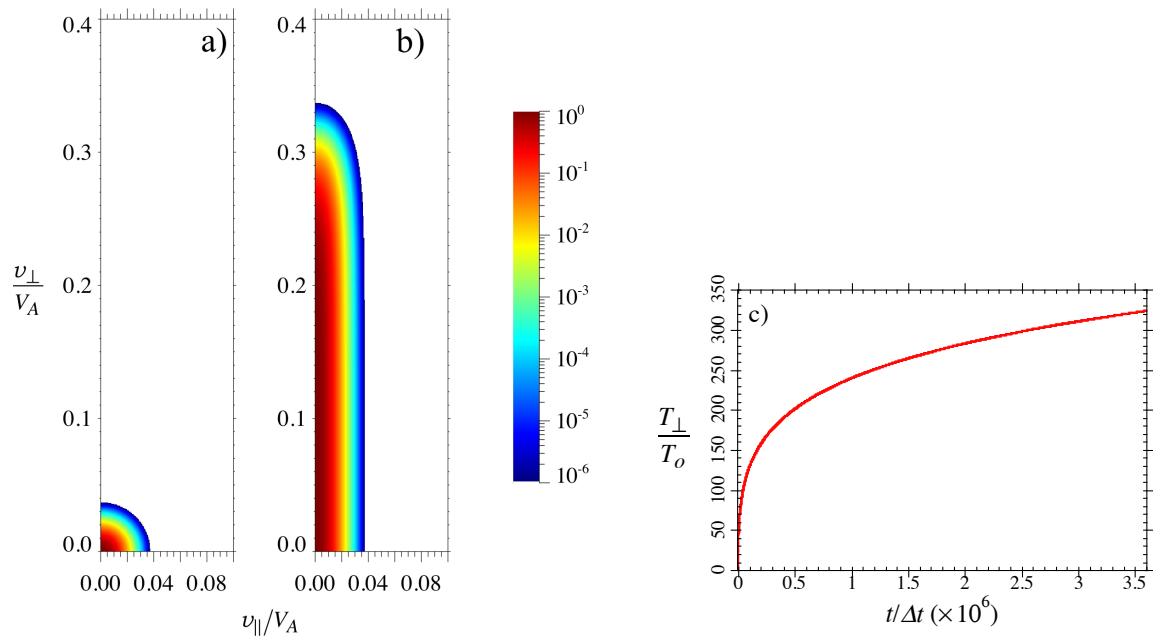

**Figure 1.** Proton distributions for MMB-only computation of §2a. Phase-space densities, each normalized to their maximum values at the origin. a) Initial distribution. b) Distribution after $3.6 \times 10^6$ time steps. c) Proton perpendicular temperature as function of time.

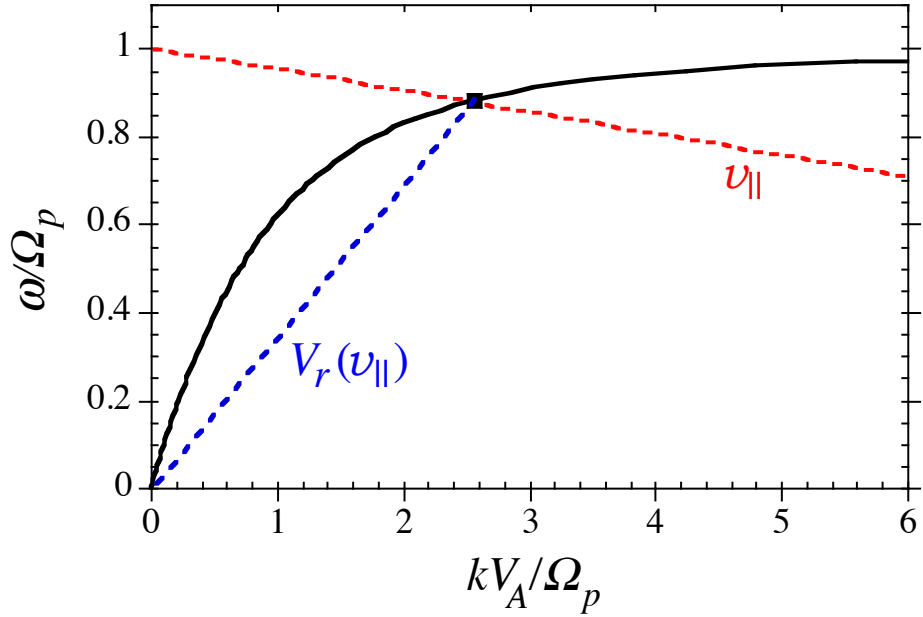

**Figure 2.** Schematic of the resonant cyclotron interaction between protons and parallel-propagating IC waves. The black line is the wave dispersion relation (10) for a cold electron-proton plasma. The red dashed line is an example of the cyclotron resonance condition (8) for a proton with $v_\| < 0$. The slope of the blue dashed line, $V_r (v_\|)$, gives the phase speed for the corresponding IC wave resonant with that proton. It is clear, by varying the proton parallel speed and considering the coupled variation of the resonant phase speed, that the resonant surfaces given by (11) - (13) describe a nested family of surfaces that do not intersect one another.

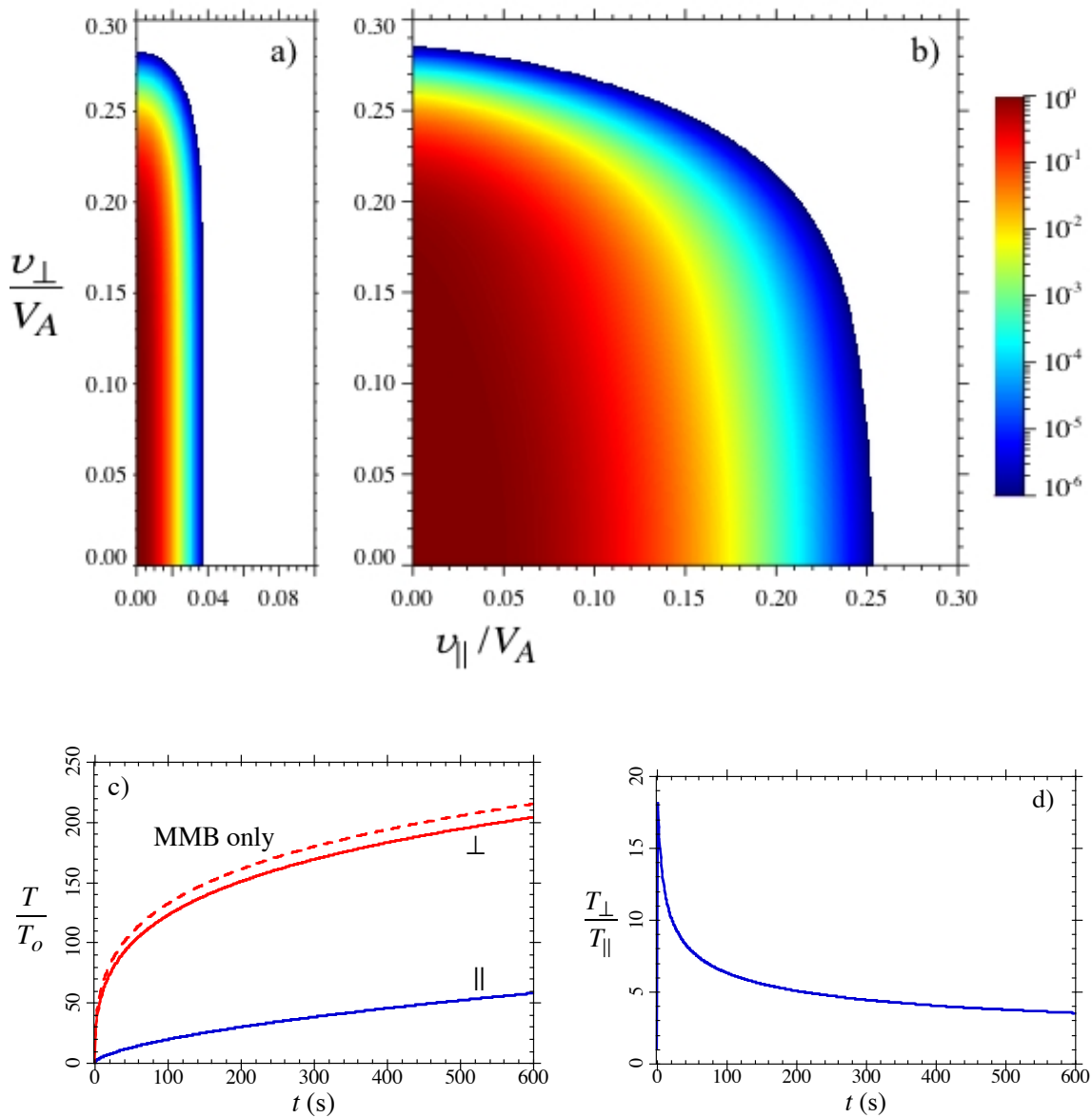

**Figure 3.** Proton distributions for Case A. Phase-space densities after 600 s, each normalized to their maximum values at the origin, for a) MMB heating only, b) combined evolution of MMB heating and quasilinear scattering. c) Parallel (solid blue line) and perpendicular (solid red line) proton temperatures as functions of time for the combined evolution, along with the perpendicular temperature for MMB heating only (red dashed line). d) Proton anisotropy for the combined evolution as a function of time.

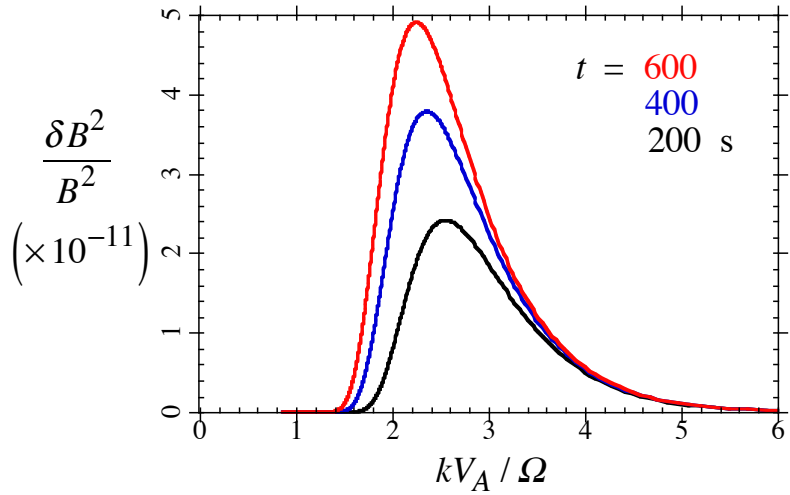

**Figure 4.** Resonant IC wave intensity spectra after 200 (black), 400 (blue) and 600 (red) seconds, for Case A.

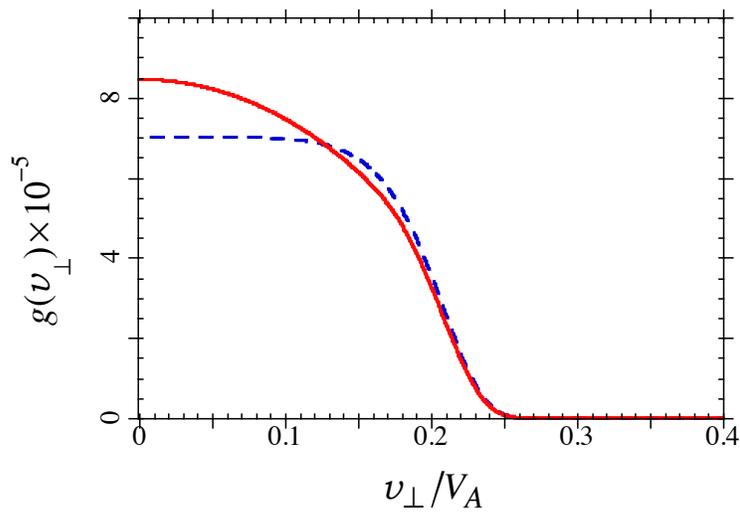

**Figure 5.** Proton reduced distributions (1) at 600 s for the MMB heating only evolution (blue dashed line) and the combined MMB and quasilinear evolution (solid red line).

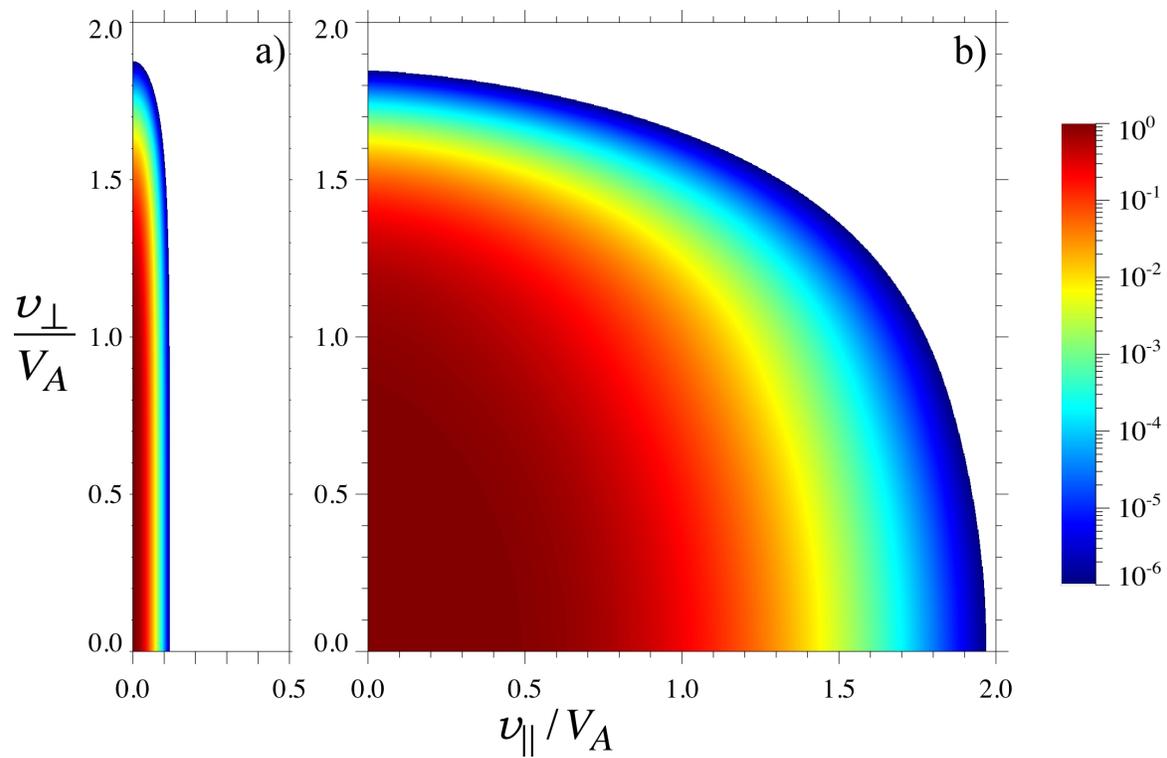

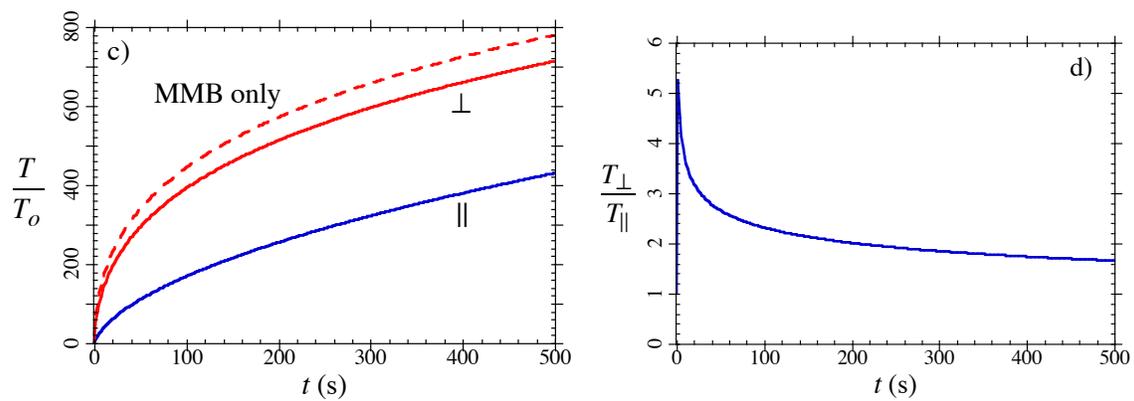

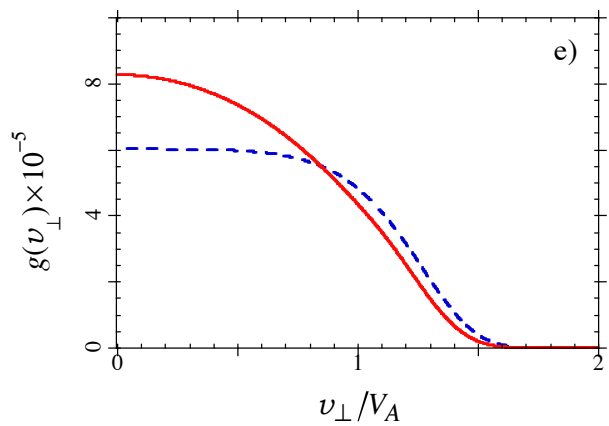

**Figure 6.** Proton distributions for Case B. Phase-space densities after 500 s, each normalized to their maximum values at the origin, for a) MMB heating only, b) combined evolution of MMB heating and quasilinear scattering. c) Parallel (solid blue line) and perpendicular (solid red line) proton temperatures as functions of time for the combined evolution, along with the perpendicular temperature for MMB heating only (red dashed line). d) Proton anisotropy for the combined evolution as a function of time. e) Reduced distributions at 500 s for the combined evolution (solid red line) and the MMB heating only (blue dashed line).

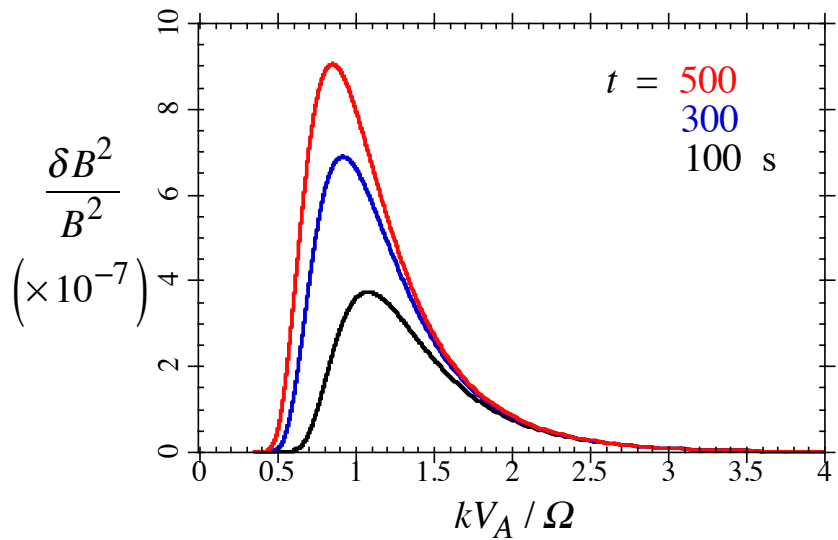

**Figure 7.** Resonant IC wave intensity spectra after 100 (black), 300 (blue) and 500 (red) seconds, for Case B.

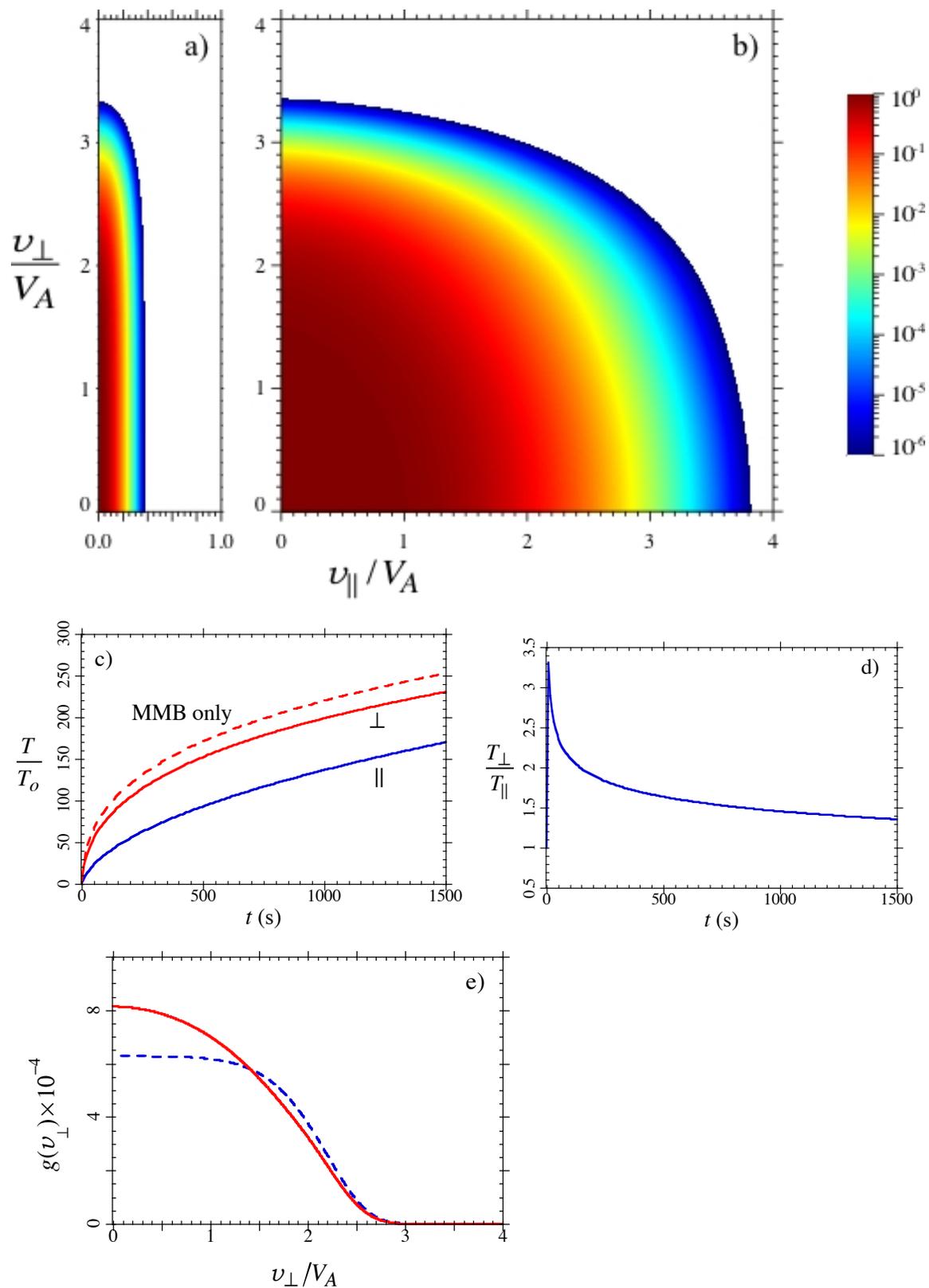

**Figure 8.** Proton distributions as in Figure 6, at 1500 s for Case C.

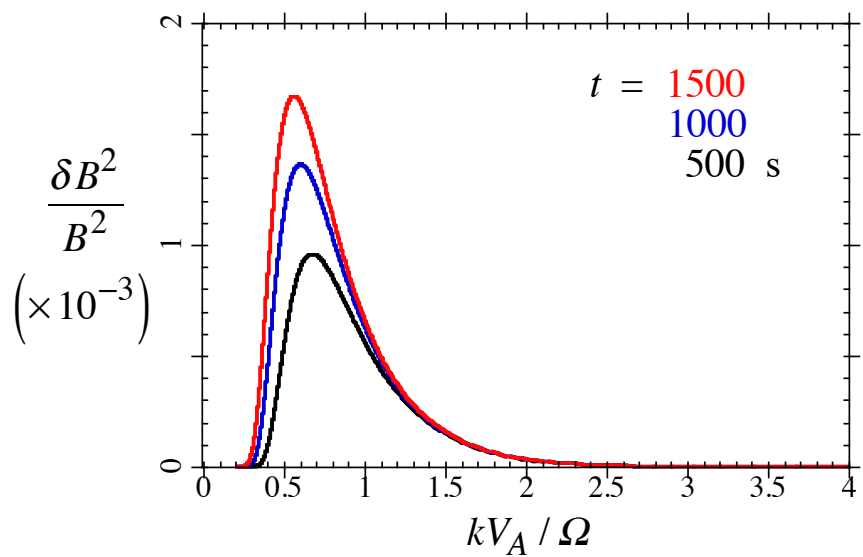

**Figure 9.** Resonant IC wave intensity spectra after 500 (black), 1000 (blue) and 1500 (red) seconds, for Case C.

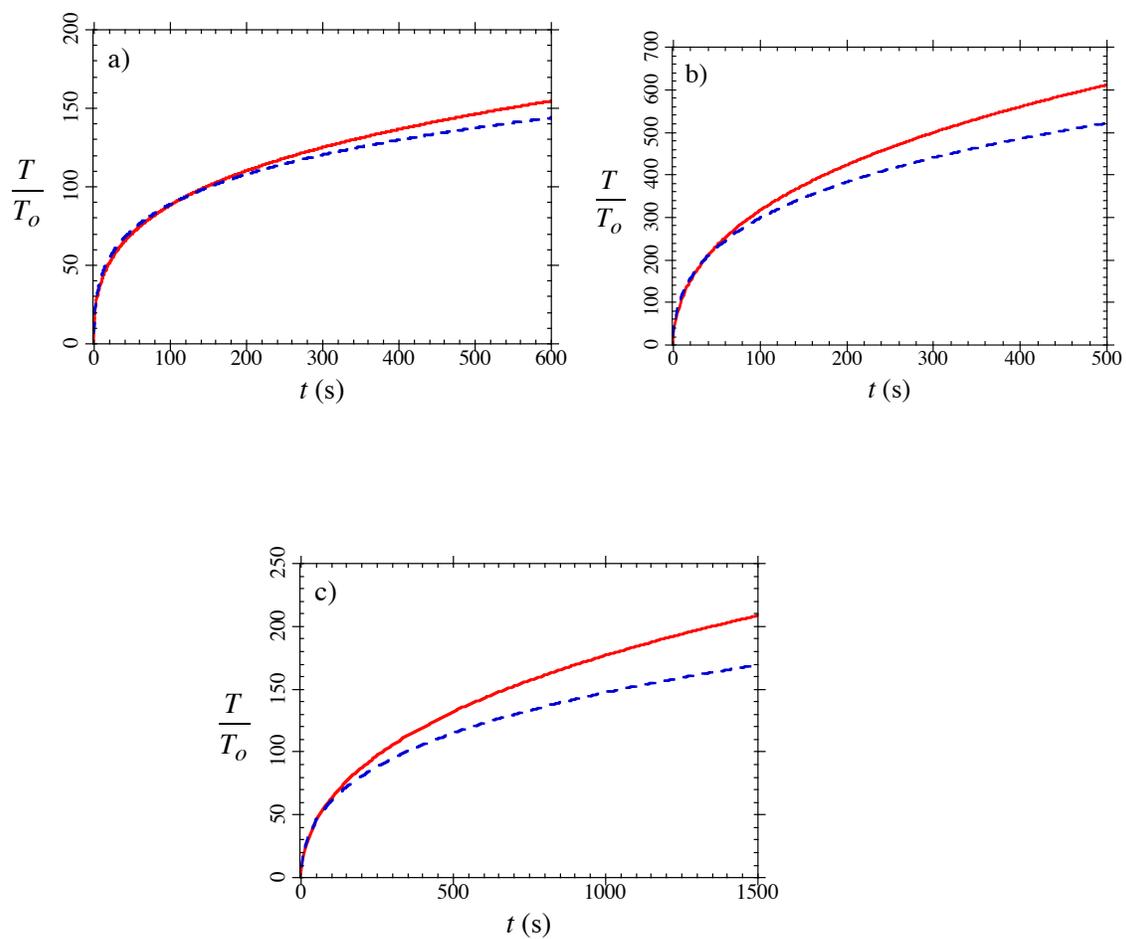

**Figure 10.** Total proton temperatures, $T = (2\,T_\perp + T_\parallel)$, for the combined evolution (solid red line) and MMB heating only (dashed blue line) in a) Case A, b) Case B, and c) Case C.